\documentclass{emulateapj}
\usepackage{natbib}
\usepackage{apjfonts}
\usepackage{multirow}

\newcommand{\rl}{$R_{\rm BLR} - L$}
\newcommand{\msigma}{$M_{\rm BH}-\sigma_{\star}$}

\newcommand{\mbh}{$M_{\rm BH}$}

\newcommand{\actaa}{Acta Astronomica}

\newcommand{\msun}{$M_{\odot}$}

\shorttitle{Reverberation Mapping of NGC\,3783}
\shortauthors{Bentz et al.}

\received{}
\accepted{}

\begin{document}

\title{Robotic Reverberation Mapping of the Southern Seyfert NGC\,3783}

\author{ Misty~C.~Bentz\altaffilmark{1},
Rachel~Street\altaffilmark{2},
Christopher~A.~Onken\altaffilmark{3},
Monica~Valluri\altaffilmark{4}
}

\altaffiltext{1}{Department of Physics and Astronomy,
		 Georgia State University,
		 Atlanta, GA 30303, USA;
		 bentz@astro.gsu.edu}
\altaffiltext{2}{LCOGT, 6740 Cortona Drive, Suite 102, 
        Goleta, CA 93117, USA}		 
\altaffiltext{3}{Research School of Astronomy and Astrophysics, 
        Australian National University, 
        Canberra, ACT 2611, Australia }
\altaffiltext{4}{Department of Astronomy,
         University of Michigan,
         Ann Arbor, MI, 48104, USA}

\begin{abstract}
We present spectroscopic and photometric monitoring of NGC\,3783 conducted throughout the first half of 2020.  Time delays between the continuum variations and the response of the broad optical emission lines were clearly detected, and we report reverberation measurements for H$\beta$, \ion{He}{2} $\lambda 4686$, H$\gamma$, and H$\delta$.  From the time delay in the broad H$\beta$ emission line and the line width in the variable portion of the spectrum, we derive a black hole mass of $M_{\rm BH} = 2.34^{+0.43}_{-0.43} \times 10^7$\,\msun.  This is slightly smaller than, but consistent with, previous determinations.  However, our significantly improved time sampling ($T_{\rm med}=1.7$\,days compared to $T_{\rm med}=4.0$\,days) has reduced the uncertainties on both the time delay and the derived mass by $\sim 50$\%.  We also detect clear velocity-resolved time delays across the broad H$\beta$ profile, with shorter lags in the line wings and a longer lag in the line core.  Future modeling of the full velocity-resolved time delay response will further improve the reverberation-based mass for NGC\,3783, adding it to the small but growing sample of AGNs for which we have constrained the black hole mass as well as the geometry and kinematics of the broad line region. Upcoming MUSE observations at VLT will also allow NGC\,3783 to join the smaller sample of black holes where reverberation masses and masses from stellar dynamical modeling may be directly compared.

\end{abstract}

\keywords{galaxies: active --- galaxies: nuclei --- galaxies: Seyfert}

\section{Introduction}

Over the last $\sim$25 years, many studies have led to the understanding that supermassive black holes ($M_{\rm BH} = 10^6-10^{10}$\,M$_{\odot}$) play a significant role in galaxy evolution and cosmology.  The evidence comes from both computational modeling (e.g., \citealt{kauffmann00,granato04,dimatteo05,springel05,hopkins06,ciotti10,scannapieco12}) and observational results (e.g., \citealt{magorrian98,ferrarese00,gebhardt00,gultekin09,kormendy13,vandenbosch16,bentz18}).
Yet, while it is now clear that black holes play an important role in the growth of structure throughout the evolution of our universe, and that this role likely involves feedback from active periods of accretion in the black hole's life, the exact nature and physical manifestations of the symbiotic relationship between galaxies and black holes are not well understood.  One key piece to unraveling this mystery rests on an accurate determination of the mass of the central black hole.

Currently, there are only a few methods that are able to directly constrain the gravitational influence of the invisible black hole on luminous tracers (stars or gas) and thus measure \mbh.  In our own galaxy, this has been accomplished with long-term astrometric monitoring of individual stars in the Galactic Center \citep{ghez00,genzel00,ghez08}.  For nearby galaxies ($D \lesssim 100$\,Mpc), however, individual stars cannot be resolved in the galactic nuclei, so the most widely-used methods instead involve dynamical modeling of the bulk motions of stars or gas (e.g., \citealt{macchetto97,vandermarel98,barth01,gebhardt03}).  These methods are constrained by the spatial resolution achievable with the current generation of large telescopes, and are thus inherently limited by the distances to the galaxies.  In active galaxies, on the other hand, black hole masses are most often derived from reverberation mapping, in which light echoes within the photoionized gas around the black hole are used to measure physical distances that are otherwise spatially unresolvable \citep{blandford82,peterson93}.

Both the dynamical and reverberation techniques currently suffer from several inherent uncertainties and potential systematic biases (cf.\ \citealt{peterson10,graham11,kormendy13}).  Both techniques require certain criteria to be fulfilled before they may be applied, which means that, practically speaking, there are few galaxies where $M_{\rm BH}$ may be directly constrained through multiple independent techniques to directly test their consistencies. This is particularly difficult for comparisons of reverberation and stellar dynamical masses: active galaxies with broad emission lines are rare in the local universe, and so almost all galaxies where reverberation mapping may be applied are too distant to achieve the spatial resolution needed to probe the gravitational influence of the black hole on the stellar dynamics.

Dynamical and reverberation mass measurements are critical to our current understanding of galaxy and black hole growth and co-evolution.  The black hole scaling relationships that are derived from these direct measurements are fundamental for observational studies of black hole mass/luminosity functions as well as cosmological simulations of galaxy evolution.  And yet we do not know if dynamical modeling and reverberation mapping give consistent black hole masses when applied to the same galaxies.  The recent highly-publicized Event Horizon Telescope results underscored the importance of carrying out black hole mass comparisons by demonstrating that in the case of M87, the black hole mass derived from interferometry agrees with the stellar dynamical mass but not the gas dynamical mass \citep{eht19}.

To date, only two AGNs (NGC\,4151 and NGC\,3227) have $M_{\rm BH}$ derived from both stellar dynamical modeling and reverberation mapping \citep{davies06,bentz06b,denney10,onken14}.  The mass measurements for NGC\,3227 cover a range of approximately an order of magnitude, with the dynamical masses consistently higher than the reverberation masses.  On the other hand, the masses for NGC\,4151 show reasonable agreement with each other.  As one of the nearest ($D\approx40$\,Mpc) and apparently brightest broad-lined Seyferts, NGC\,3783 is one of the best candidates for carrying out these important consistency checks.

While NGC\,3783 is an obvious candidate for such a study, it currently lacks an H$\beta$ reverberation measurement with the accuracy needed for mass comparisons. NGC\,3783 was one of the very first AGNs to be studied through reverberation mapping.  This early attempt \citep{stirpe94} employed an image-tube as the spectrograph detector, rather than a CCD, and it suffered from linearity problems.  Furthermore, the temporal cadence of the spectroscopy was quite low ($\sim$once per week) because the broad line region size, and therefore the expected time delay, was not yet well understood.  While reprocessing of the initial data set by \citet{onken02} improved the quality of the reverberation measurements, the median time sampling of 4.0\,days and a few large gaps in coverage leading to an average time sampling of 6.2\,days limited the accuracy that could be obtained.  Reanalysis of the light curves by \citet{zu11} resulted in significantly smaller uncertainties, but ultimately this relies on several assumptions regarding the underlying behavior of the light curves.

We therefore conducted a new monitoring campaign to constrain the optical reverberations in NGC\,3783.  Our first attempt, using the CTIO 1.5\,m telescope in 2012, was unsuccessful because of an extended gap in time coverage in the middle of the campaign due to a scheduling mistake.  Our second attempt in 2017 was also unsuccessful because too few spectra were acquired before the observing season for the target had ended.  
In this paper, we describe the initial results from our third, and finally successful, attempt.

\section{Observations}

NGC\,3783 is an active barred spiral galaxy located at $\alpha=$11:39:01.7 and $\delta=-$37:44:19 with a redshift of $z=0.0097$.  NGC\,3783 was monitored photometrically and spectroscopically throughout Semester 2020A with the Las Cumbres Observatory global telescope (LCOGT) network in the Southern hemisphere (NOAO 2020A-011, PI: Bentz).

\subsection{Imaging}

$V-$band photometric monitoring was requested with the Sinistro cameras on the LCOGT network of robotic 1-m telescopes.  Images were scheduled to be acquired every $\sim8$\,hours beginning in mid-February, but weather and an unexpected and unprecedented global pandemic reduced the time sampling considerably.

Observations began on 12 February 2020 and were initially carried out at Cerro Tololo Inter-American Observatory (CTIO),  South African Astronomical Observatory (SAAO), and Siding Spring Observatory (SSO).  However, observations were halted at CTIO on 18 March 2020 and at SAAO on 26 March 2020 due to the global spread of the novel coronavirus.  While the LCOGT telescopes are robotic, they do require regular maintenance and troubleshooting. CTIO and SAAO, along with most of the major observatories around the world, temporarily suspended all of their mountaintop operations to protect the health and safety of their staff.  On 6 May 2020, observations were resumed at SAAO.  Observations at SSO continued uninterrupted throughout the semester, and the monitoring program finished as planned at the end of the observing semester on 30 June 2020. 

Over the course of the semester, 209 $V-$band images were acquired: 100 from SSO, 68 from SAAO, and 41 from CTIO. 
Exposures were set up identically across the observatories, each with an exposure time of 60\,s and acquired at a typical airmass of 1.09.  The Sinistro cameras have a field of view of $26\farcm5\times26\farcm5$ and an angular sampling of 0.389\arcsec/pixel.  The LCOGT pipeline applies typical bias, dark, and flat field corrections to the raw images and serves up fully reduced images in the archive.

\begin{figure}
    \centering
    \epsscale{1.17}
    \plotone{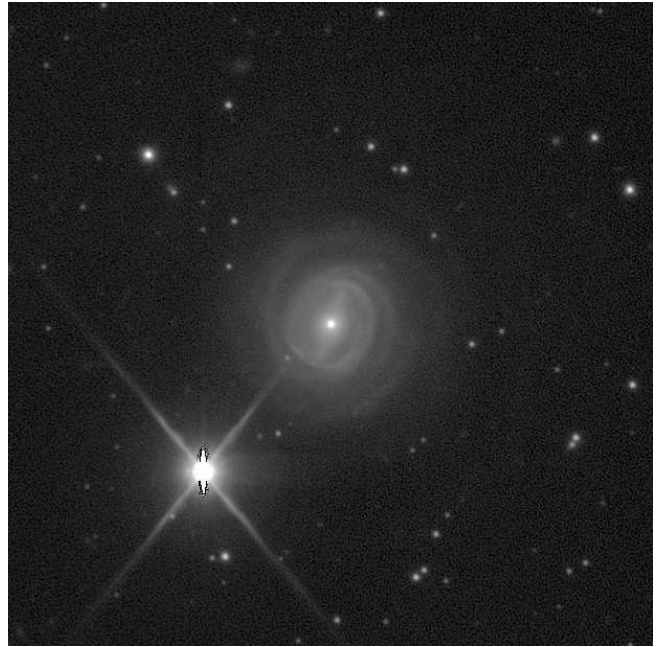}
    \caption{A portion of the reference image of NGC\,3783 built from the best individual $V-$band frames collected throughout the monitoring program.  The central $4\arcmin \times 4\arcmin$ are shown, oriented with North up and East to the left.}
    \label{fig:refimg}
\end{figure}

After downloading the reduced images from the archive, we registered all of them to a common reference frame using the {\sc Sexterp} package \citep{siverd12}.  We then employed the image subtraction package {\sc ISIS} \citep{alard00,alard98} to build a reference frame from the best images, convolve the reference frame to match the characteristics of each of the 209 individual images, and then subtract the convolved reference from each image to remove all non-varying sources.  The residual flux in the nucleus of NGC\,3783, which is relative to the brightness of the AGN in the reference image (see Figure~\ref{fig:refimg}) and may therefore appear as either positive or negative counts, was then measured with aperture photometry.  

To convert the residual nuclear flux from counts to calibrated photometry, we modeled the reference frame with {\sc Galfit} \citep{peng02,peng10}.  By constraining the host-galaxy surface brightness features with analytical models, we were able to isolate the contribution of the central AGN in the reference image.  Using $V-$band measurements of several field stars from the AAVSO Photometric All Sky Survey catalog  \citep{henden14}, we determined the magnitude zeropoint of the reference image, thus defining the conversion from counts to calibrated magnitudes or fluxes for the reference brightness of the AGN.  With the reference brightness constrained, the residual fluxes were then calibrated.  

Comparison of the measurements from the three observing sites demonstrated that there were slight photometric offsets (on the order of $\sim 0.05-0.10$\,mag) between the images from SSO, SAAO, and CTIO.  The majority of our measurements came from SSO, so we adopted that data set as the photometric anchor. We then determined the best linear fit between measurements from SAAO and SSO that were closely spaced in time ($\lesssim 0.5$\,day separation), and between closely spaced measurements from CTIO and SSO.  Measurements from SAAO and CTIO were then scaled by these best-fit linear relationships to match the SSO measurements.  

\begin{figure}
    \epsscale{1.15}
    \plotone{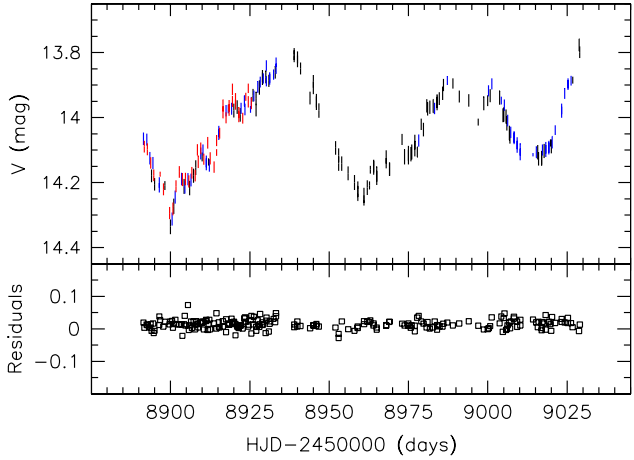}
    \caption{Top: $V-$band photometric light curve of the nuclear emission in NGC\,3783. Black points were observed at SSO, blue points were observed at SAAO, and red points were observed at CTIO. Bottom: Magnitude residuals after image subtraction for a $V=14$\,mag non-varying isolated field star.}
    \label{fig:vlc}
\end{figure}

Finally, the uncertainties on the photometry from image subtraction are known to be underestimated in many cases (e.g., \citealt{zebrun01,hartman05}).  We examined the residual counts for non-varying field stars in the subtracted images, and determined the scale factor for the uncertainties that was needed to account for this additional scatter. Our procedure closely followed that of \citet{hartman04}.  The resulting $V-$band light curve is displayed in Figure~\ref{fig:vlc}. 

Given the strong variability in the nucleus of NGC\,3783 over the course of the monitoring program, differential photometry would produce a similar $V-$band light curve.  However the variations would be damped by the host-galaxy flux contribution, and additional noise may be introduced through seeing variations that affect the amount of starlight in the photometric aperture.  By removing the intrinsically non-variable host galaxy through image subtraction, we recover the most accurate measurements of the continuum variations that are possible with these data.

\subsection{Spectroscopy}

\begin{figure}
    \epsscale{1.15} 
    \plotone{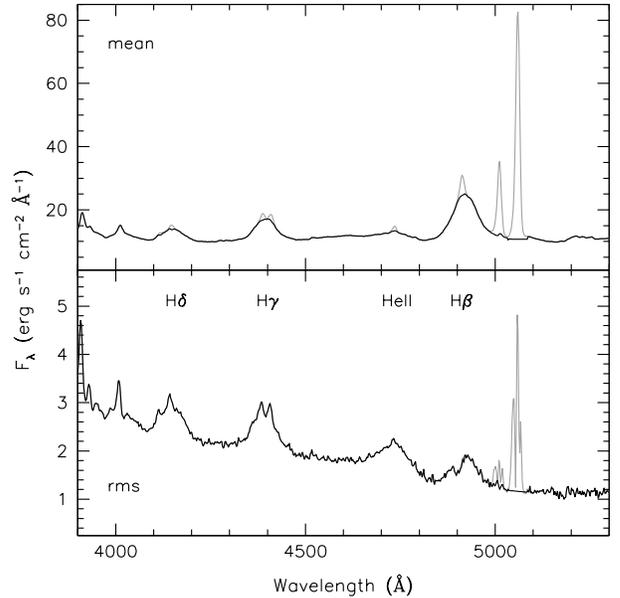}
    \caption{Mean ({\it upper}) and root-mean-square ({\it lower}) of all the blue spectra of NGC\,3783 collected throughout the monitoring campaign. Text labels mark the wavelengths for the main broad emission lines in the mean spectrum, with their variable components visible in the rms spectrum.  Narrow line subtracted spectra are displayed with the black line, while the contributions from several narrow emission lines are displayed in gray.}
    \label{fig:meanrms}
\end{figure}

Spectra were acquired with the robotic FLOYDS spectrograph on the 2-m
Faulkes Telescope South at SSO.  FLOYDS is a low-resolution
cross-dispersed spectrograph that covers 540-1000\,nm in the first
order and 320-570\,nm in the second order.  To minimize the effects of
variable seeing on the spectrophotometry, we employed the
6\arcsec\ slit rotated to a fixed position angle of 0\degr\ (N-S
orientation) on the sky for all observations.  We specifically did
  not align the slit to the parallactic angle because the strong bar
  in the host galaxy would then induce a variable host-galaxy
  contribution in the spectra, making intrinsic nuclear variability
  more difficult to detect.  We limited our observations to airmasses
  <1.6 so that the use of the 6\arcsec\ slit would avoid most
  potential slit losses due to differential refraction at the bluest
  wavelengths investigated here.

Spectra were scheduled to be acquired every $\sim24$\,hours, although
weather and oversubscription of the telescope reduced the actual time
sampling that was achieved.  A total of 50 spectra were obtained, each
with an exposure time of 900\,s and at a typical airmass of 1.26.
The median seeing value during the observations was 2.2\arcsec.
Each observing sequence included HgZn arc lamp images that were taken
immediately before and after the science spectrum, as well as a flat
field image.

The LCOGT reduction pipeline applies typical bias, dark, and flat field corrections, rectifies the two orders of the spectra, and applies rough wavelength and flux calibrations based on a historical wavelength solution and sensitivity function.  We downloaded the rectified two-dimensional arc lamp spectra and science spectra from the archive and began our custom reductions with those files. Working in IRAF\footnote{IRAF is distributed by the National Optical Astronomy Observatories, which are operated by the Association of Universities for Research in Astronomy, Inc., under cooperative agreement with the National Science Foundation.}, we cleaned the science spectra of cosmic rays and then extracted one-dimensional science spectra and arc lamp spectra with an extraction width of 10\,pixels.  The FLOYDS camera has a pixel scale of 0.337\arcsec/pixel, so the extraction width corresponds to an angular size of 3.37\arcsec.  We then manually identified the arc lamp lines for all 50 visits and applied the updated wavelength solutions to the corresponding science spectra.  

We then corrected for small differences between the nightly spectra, including small wavelength shifts, offsets in the flux calibration, and differences in resolution from seeing variations.  We applied the \citet{vangroningen92} scaling algorithm, focusing on the  [\ion{O}{3}] doublet region of the spectra.  Each spectrum is compared to a reference spectrum, generally a combination of the best spectra as identified by the user, and the algorithm applies small shifts and smoothing to minimize the differences between the two. In this way, the [\ion{O}{3}] emission lines are treated as internal flux calibration sources that are nonvariable on the timescales probed here (cf.\ \citealt{peterson13}).  Spectra that are treated in this way have been shown by \citet{peterson98a} to have relative spectrophotometry that is accurate to $\sim 2$\% throughout the duration of the monitoring program. Without a similarly strong and unblended emission line in the red spectra, we are unable to properly calibrate the H$\alpha$ emission region, and so we focused the analysis solely on the blue spectra. 
 
We checked the accuracy of the spectral scaling method by examining the curves of growth for [\ion{O}{3}] $\lambda$5007, which may be expected to arise from a region that is marginally resolved spatially, and for H$\beta$, which is dominated by broad emission that arises from a spatially unresolved region. We find that the curves of growth are the same for [\ion{O}{3}] and H$\beta$ under the range of seeing conditions encountered in the monitoring program, with the largest differences occurring at the 1.7\% level.  Thus the assumption holds that seeing variations affect measurements of [\ion{O}{3}]  flux in the same way as the H$\beta$ flux, and that these spectra may be intercalibrated at an accuracy that is better than 2\%  using the [\ion{O}{3}] emission lines.

The rough flux calibration determined from the historical sensitivity function does not provide accurate absolute spectrophotometry, however.  The typical recommendation for observers is to search the LCOGT archive for standard star spectra that were acquired on the same night as the target spectra, however, the standard star spectra are acquired through a 2\arcsec\ slit by default and are therefore not suitable for fully calibrating our target spectra, which were acquired through the 6\arcsec\ slit.  Furthermore, measured values of the integrated [\ion{O}{3}] $\lambda 5007$\,\AA\ emission line flux vary over almost an order of magnitude in the literature, from $3.8\times10^{-13}$\,erg\,s$^{-1}$\,cm$^{-2}$ \citep{dopita15} on the low end to $1.4\times10^{-12}$\,erg\,s$^{-1}$\,cm$^{-2}$  \citep{ueda15} on the high end, even though [\ion{O}{3}] fluxes only vary slowly on timescales of years to decades in local Seyferts (e.g, \citealt{peterson13}).  A STIS G430M spectrum of NGC\,3783 acquired through the 0\farcs2 slit (GO-12212, PI: Crenshaw) provides an integrated flux of $7.4\times10^{-13}$\,erg\,s$^{-1}$\,cm$^{-2}$.  While [\ion{O}{3}] imaging of NGC\,3783 does show that the emission is quite compact and centered on the nucleus, \citet{fischer13} found that faint [\ion{O}{3}] emission is still clearly detected out to a radius of $\sim2\arcsec$, and so the flux determined from the STIS spectrum should be taken as a lower limit given the very narrow slit that was employed.

We had previously attempted to carry out this program in early 2017
during a special call for NOAO science to be conducted with LCOGT
facilities (NOAO 2017B-0042, PI: Bentz).  Scheduling priorities on the
2-m telescope combined with somewhat limited nightly visibility of
NGC\,3783 based on the time of year led to very poor temporal sampling
that did not allow any time delays to be detected.  However, as part
of that program we had requested observations of the
spectrophotometric standard LTT\,4364 through the 6\arcsec\ slit along
with the spectra of NGC\,3783, and so we were able to make use of
those observations to accurately constrain the integrated flux of the
[\ion{O}{3}] emission.  Starting with the rectified two-dimensional
spectra from the pipeline, which have a rough flux calibration from
the historical sensitivity function, we extracted the one-dimensional
spectra of LTT\,4364 and NGC\,3783 with the same aperture as above.
We then fit a low order polynomial to the ratio of each observed
LTT\,4364 spectrum relative to the calibrated spectrum of LTT\,4364
from \citet{bessell99}.  The low order polynomial was then used to
correct the shape of the spectrum of NGC\,3783 acquired on the same
night.  Of the 18 nights on which spectra of both targets had been
acquired, several were obviously nonphotometric based on the poor
quality of the spectra, and so they were discarded.  For the remaining
10 nights, we measured the [\ion{O}{3}] $\lambda 5007$\,\AA\ flux from
the corrected spectra by fitting a local linear continuum under the
emission line and integrating the flux above the continuum.  One night
gave a measurement that was clearly discrepant from the others, so it
was discarded.  The median of the measurements from the remaining nine
nights is $F_{5007}=(10.05\pm0.68) \times
10^{-13}$\,erg\,s$^{-1}$\,cm$^{-2}$, which we adopt as the integrated
flux of the [\ion{O}{3}] $\lambda 5007$\,\AA\ emission line.  The
final step in processing the 2020 spectra was to scale them to match
the adopted [\ion{O}{3}] $\lambda 5007$\,\AA\ flux to ensure accuracy
in the absolute spectrophotometry.  While this step is not
  important for the reverberation analysis since it does not affect
  the time delays or emission line widths, it is vital for determining
  the AGN luminosity when investigating black hole scaling
  relationships.

Figure~\ref{fig:meanrms} shows the mean and root-mean-square (rms) of the scaled spectra.  The rms spectrum clearly displays strong variability in the broad Balmer emission lines (H$\beta$, H$\gamma$, and H$\delta$) as well as the \ion{He}{2} $\lambda4686$\,\AA\ line.  There is no discernible variability from \ion{Fe}{2}, which appears as a weak emission component in the mean spectrum.

\begin{deluxetable*}{lccccccc}[h]
\tablecolumns{8}
\tablewidth{0pt}
\tablecaption{Light Curve Statistics}
\tablehead{
\colhead{Time Series} &
\colhead{N} &
\colhead{$\langle T \rangle$ (days)} &
\colhead{$T_{med}$ (days)} &
\colhead{$\langle F \rangle$} &
\colhead{$\langle \sigma_F / F \rangle$} &
\colhead{$F_{var}$} &
\colhead{$R_{max}$}\\
\colhead{(1)} &
\colhead{(2)} &
\colhead{(3)} &
\colhead{(4)} &
\colhead{(5)} &
\colhead{(6)} &
\colhead{(7)} &
\colhead{(8)} 
}
\startdata
V           & 209 & $0.7\pm0.8$ & 0.4 & ~$9.6\pm0.9$  & 0.014 & 0.092 & $1.536\pm0.034$ \\
5100\,\AA   & 50  & $2.4\pm2.1$ & 1.7 & ~$9.7\pm1.1$  & 0.020 & 0.117 & $1.692\pm0.069$ \\
H$\beta$    & 50  & $2.4\pm2.1$ & 1.7 & $11.6\pm0.7$  & 0.010 & 0.063 & $1.255\pm0.012$ \\
\ion{He}{2} & 50  & $2.4\pm2.1$ & 1.7 & ~$1.5\pm0.6$  & 0.085 & 0.405 & $4.539\pm0.777$ \\
H$\gamma$   & 50  & $2.4\pm2.1$ & 1.7 & ~$4.8\pm0.7$  & 0.020 & 0.142 & $1.802\pm0.054$ \\
H$\delta$   & 50  & $2.4\pm2.1$ & 1.7 & ~$2.4\pm0.5$  & 0.041 & 0.215 & $2.680\pm0.635$ 
\label{tab:lcstats}
\enddata 

\tablecomments{Continuum flux densities ($V$ and 5100\,\AA) are quoted in units of $10^{-15}$\,erg\,s$^{-1}$\,cm$^{-2}$\,\AA$^{-1}$ while emission-line fluxes are quoted in units of $10^{-13}$\,erg\,s$^{-1}$\,cm$^{-2}$.}
\end{deluxetable*}

\section{Light Curve Analysis}

\begin{figure}
    \epsscale{1.15} 
    \plotone{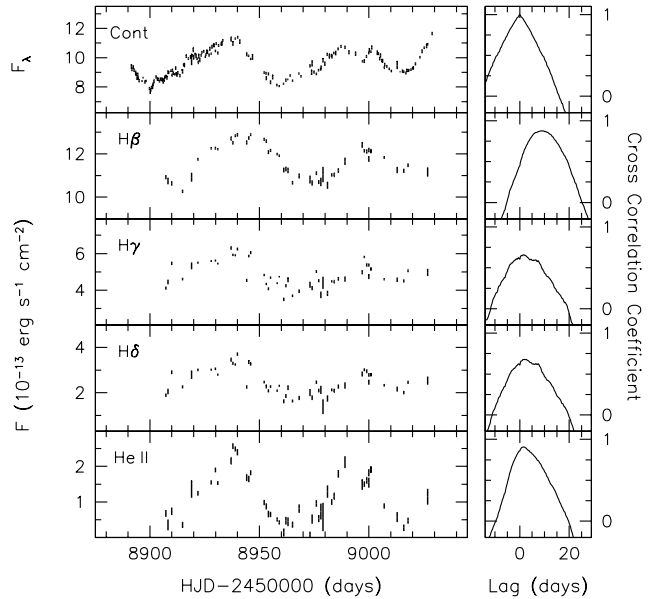}
    \caption{The merged continuum light curve and broad emission-line light curves are displayed in the left panels.  The units for the continuum flux density, F$_{\lambda}$ are $10^{-15}$\,erg\,s$^{-1}$\,cm$^{-2}$\,\AA$^{-1}$. On the right are the cross-correlation functions relative to the continuum.  For the continuum light curve, this is the auto-correlation function. All of the broad emission lines show clear time lags, with H$\beta$ exhibiting the longest lag and \ion{He}{2} the shortest. }
    \label{fig:lcall}
\end{figure}

Emission-line light curves were derived from the scaled spectra by fitting a local, linear continuum underneath each emission line of interest, and integrating the flux above the continuum.  The total flux for each emission line includes the narrow emission-line component, which is simply a constant flux offset.  We also measured the continuum flux at $5100\times(1+z)$\,\AA, a region of the spectrum that is free of emission lines and where the host-galaxy contribution is at a local minima. 

The $V-$band and $5100\times(1+z)$\,\AA\ light curves show a very similar shape, but cover slightly different portions of the time baseline.  In particular, the gap in photometric coverage between HJD$-2450000\approx 8933-8939$\,days is partially covered by the $5100\times(1+z)$\,\AA\ light curve.  We therefore scaled the $V-$band light curve to match the $5100\times(1+z)$\,\AA\ light curve, based on the linear function that described the relationship between pairs of points that were observed close together in time, and merged them together.  In our final continuum light curve, the measurements are binned with 0.25\,day bins.  All of the light curves that were used for the following analysis are displayed in Figure~\ref{fig:lcall}.

Common statistics for the light curves are tabulated in Table~\ref{tab:lcstats}.  We list the statistics for the $V-$band and $5100\times(1+z)$\,\AA\ light curves separately here, in order to be complete.  Column (1) lists the spectral feature, column (2) gives the number of measurements in the light curve, and columns (3) and (4) list the average and median time separation between measurements, respectively. Column (5) gives the mean flux and standard deviation of the light curve, while column(6) lists the mean fractional error. Column (7) lists the excess variance, computed as
\begin{equation}
    F_{var} = \frac{\sqrt{\sigma^2 - \delta^2}}{\langle F \rangle}
\end{equation}
\noindent where $\sigma^2$ is the variance of the fluxes, $\delta^2$ is their mean-square uncertainty, and $\langle F \rangle$ is the mean flux.  Column (8) is the ratio of the maximum to the minimum flux in the light curve.

The variability in the continuum light curve is clearly echoed in the broad emission line light curves.  To quantify the time delays between the variations in the continuum light curve and the variations in an emission-line, we first employed the interpolated cross-correlation function (ICCF) method of \citet{gaskell86} and \citet{gaskell87} with the modifications of \citet{white94}.  The ICCF method determines the cross-correlation function (CCF) by averaging together the two CCFs that are calculated when the continuum light curve is interpolated and then when the emission-line light curve is interpolated.  The CCFs are displayed in the panels on the right side of Figure~\ref{fig:lcall}.

The CCFs may be characterized by the peak value ($r_{\rm max}$), the time delay at which the peak occurs ($\tau_{\rm peak}$), and the centroid of the CCF ($\tau_{\rm cent}$) near the peak above some value, usually $0.8 r_{\rm max}$.  To quantify the uncertainties on $\tau_{\rm cent}$ and $\tau_{\rm peak}$, we employ the Monte Carlo flux randomization/random subset sampling (FR/RSS) method \citep{peterson98b,peterson04}.  The random subset sampling accounts for the effects of including/excluding any particular data points in the light curves.  From a light curve with $N$ data points,  $N$ points are selected without regard to whether a point has already been selected.  For a point that is selected $1\leq  n \leq N$ times, the uncertainty on that  point is scaled by a factor of $n^{1/2}$, while the typical number of points that are not selected in any specific realization is $\sim 1/e$.  The flux randomization then takes the newly sampled light curve and adjusts the flux values of the points randomly with a Gaussian deviation of the flux uncertainty.  The CCF of the randomized, sampled light curve is then calculated using the ICCF method and $r_{max}$, $\tau_{\rm peak}$, and $\tau_{\rm cent}$ are recorded.  This process is then repeated, and distributions of CCF measurements were built up over 1000 realizations.  We take the median of each distribution as the measurement value, and the uncertainties are defined to exclude the upper 15.87\% and lower 15.87\% of the realizations (corresponding to $\pm 1\sigma$ for a Gaussian distribution).  Table~\ref{tab:lagwidth} lists the observed-frame time delays measured in this way for  each broad emission line.

\begin{figure}
    \epsscale{1.15} 
    \plotone{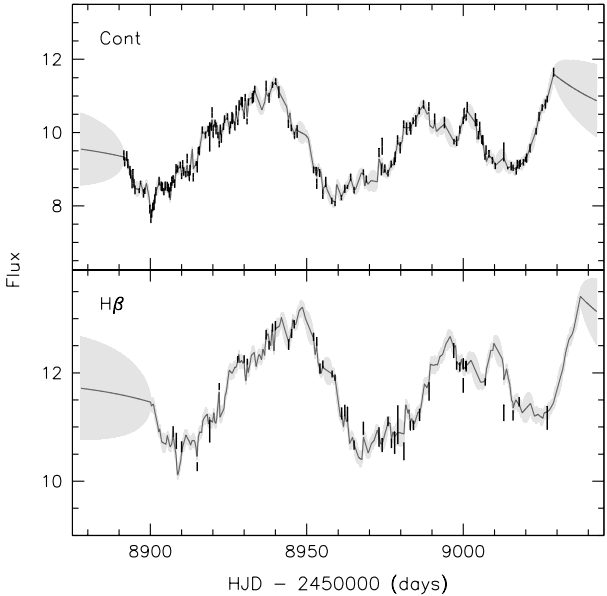}
    \caption{Continuum and H$\beta$ light curves (data points) with the mean {\tt Javelin} model light curves (solid lines) and uncertainties (gray shaded regions) overlaid.  The uncertainties on the models were derived from the standard deviation of the individual realizations. }
    \label{fig:jav}
\end{figure}

We also constrained the time delays using the {\tt Javelin} package \citep{zu11}.  {\tt Javelin} fits a damped random walk model to the continuum light curve and then determines the best top hat model for reprocessing the continuum light curve to match the emission-line light curve.  Uncertainties on the model parameters are assessed through a Bayesian Markov Chain Monte Carlo method.  {\tt Javelin} is capable of fitting multiple light curves simultaneously, but our initial experiments showed that the results were very sensitive to noise in the light curves.  We thus modeled the light curve of each emission line separately with {\tt Javelin}, as we show in Figure~\ref{fig:jav} for the H$\beta$ emission line, and we report the best-fit observed time delays as $\tau_{\rm jav}$ in Table~\ref{tab:lagwidth}.

Finally, we also attempted to constrain the time lags using a different method of treating the spectra.  Rather than carefully scaling the nightly spectra using the [\ion{O}{3}] lines, we instead measured the equivalent widths of the broad emission lines in each reduced, but unscaled, spectrum.  We then multiplied the equivalent widths by the $V-$band flux measured close in time, or interpolated between measurements when needed, to approximate the calibrated integrated flux of each emission line.  While this method has the advantage of not requiring the presence of strong isolated narrow lines and it avoids carrying out the time intensive scaling step, it assumes that seeing and aperture effects influence the emission lines in the same way as the continuum.  For a bright, extended host galaxy like NGC\,3783, this assumption may not hold.  The resulting light curves for H$\beta$ and \ion{He}{2} are similar, but slightly more noisy, than those in Figure~\ref{fig:lcall} and result in consistent time delay measurements.  The light curves for H$\gamma$ and H$\delta$ share the same general shapes as before, but result in time delay measurements that are longer than found with the spectral scaling method.   H$\alpha$ produced a noisy light curve, and while the first peak in the continuum light curve was recovered,  the subsequent peaks and valleys were lost in the noise.  Nevertheless, the recovered time delay was consistent with that measured for H$\beta$.  In the case of H$\alpha$, in particular, the varying amount of host-galaxy flux due to seeing effects seems to have detracted from the potential benefits of this method.  Similar tests with quasars that have minimal host-galaxy contamination may be more successful.

With the new H$\beta$ time delays determined here, we examined the location of NGC\,3783 on the AGN \rl\ relationship.  Following the methods outlined in \citet{bentz09b,bentz13}, we estimated the host-galaxy contribution to the 5100\,\AA\ flux from a high-resolution Hubble Space Telescope image of NGC\,3783, finding $f_{\rm gal} = (2.76 \pm 0.28) \times 10^{-15}$\,erg\,s$^{-1}$\,cm$^{-2}$\,\AA$^{-1}$,  approximately 28\% of the continuum flux during the monitoring campaign. The luminosity distance to NGC\,3783 is $\sim 42$\,Mpc, which is well within the volume where peculiar velocities may be problematic.   \citet{kourkchi17} assigned NGC\,3783 to a group of nine galaxies, only two of which have distance measurements, and report a group-averaged distance of $42\pm6$\,Mpc.  After correcting for the starlight contribution to the continuum flux and adopting a distance of $42$\,Mpc, we find a nuclear luminosity of $\log \lambda L_{5100} = 43.02\pm0.02$\,erg\,s$^{-1}$.  The best fit to the \rl\ relationship reported by \citet{bentz13} predicts an H$\beta$ time delay of $10.1\pm1.8$\,days for the luminosity of NGC\,3783, which agrees well with the time delay reported here.

\begin{deluxetable*}{lccccccc}
\tablecolumns{8}
\tablewidth{0pt}
\tablecaption{Emission-Line Time Lags and Widths}
\tablehead{
\colhead{} &
\colhead{} &
\colhead{} &
\colhead{} &
\multicolumn{2}{c}{mean} &
\multicolumn{2}{c}{rms}\\
\colhead{Line} &
\colhead{$\tau_{\rm cent}$} &
\colhead{$\tau_{\rm peak}$} &
\colhead{$\tau_{\rm jav}$} &
\colhead{FWHM} &
\colhead{$\sigma_{\rm line}$} &
\colhead{FWHM} &
\colhead{$\sigma_{\rm line}$}\\
\colhead{} &
\colhead{(days)} &
\colhead{(days)} &
\colhead{(days)} &
\colhead{(km s$^{-1}$)} &
\colhead{(km s$^{-1}$)} &
\colhead{(km s$^{-1}$)} &
\colhead{(km s$^{-1}$)}
}
\startdata
H$\beta$    & $9.60^{+0.65}_{-0.72}$ &	$9.50^{+1.00}_{-1.00}$ & $8.64^{+1.71}_{-0.11}$ & $4486\pm35$	& $1825\pm19$  & $4728\pm676$ &	$1619\pm137$ \\
\ion{He}{2} & $1.95^{+1.02}_{-0.98}$ &	$1.50^{+0.50}_{-0.75}$ &  $1.35^{+0.10}_{-0.08}$ & \nodata      & \nodata 	   & $4771\pm638$ & $2146\pm72$   \\
H$\gamma$   & $2.66^{+1.35}_{-1.40}$ &	$2.25^{+2.00}_{-1.25}$ &  $2.46^{+2.30}_{-0.06}$ & $4304\pm79$  & $1621\pm22$  & $4148\pm394$  &	$1692\pm36$ \\
H$\delta$   & $3.39^{+1.17}_{-1.29}$ &	$3.00^{+2.00}_{-1.75}$ &  $4.76^{+0.04}_{-2.31}$ & $4274\pm100$ & $1611\pm30$  & $4035\pm461$  &	$1709\pm49$ 
\label{tab:lagwidth}
\enddata 

\tablecomments{Time lags are listed as measurements in the observer's frame.}
\end{deluxetable*}

\section{Emission Line Widths}

The width of a broad line gives an estimate of the line-of-sight velocity of gas in the broad line region. Many previous reverberation experiments have shown that the typical velocities of the broad line region (BLR) gas that responds to continuum variations, the rms line profiles, may not be the same as those measured from the integrated line emission (the mean line profiles).  Furthermore, the narrow emission is blended with the broad emission in the spectra, but is emitted from a physically distinct region with different kinematics.  For these reasons, the line widths measured in the rms spectrum are generally preferred, as they correspond only to gas that reverberates.  Furthermore, the second moment of the line profile, $\sigma_{\rm line}$, is generally preferred over the full width at half maximum, FWHM, because it is less susceptible to biasing from narrow line emission and noise (e.g., \citealt{peterson04}).

For completeness, however, we measured the broad emission line widths in both the mean and the rms spectra, and we report both FWHM and $\sigma_{\rm line}$.  Before measuring the line widths, we did attempt to subtract the narrow components from the spectra.  The [\ion{O}{3}] $\lambda 5007$ line was used as a template, and was shifted and scaled to match the narrow components of H$\beta$, \ion{He}{2}, H$\gamma$, and H$\delta$, as well as additional contaminating narrow line emission from [\ion{S}{2}] $\lambda 4071$ and [\ion{O}{3}] $\lambda 4363$.  Some residual noise from narrow emission remains in the rms spectrum at the bluest wavelengths, where the intercalibration of the nightly spectra is less accurate (cf.\ Figure~\ref{fig:meanrms}).

Line widths were measured directly from the narrow-line subtracted spectra, with a local linear continuum defined below each emission line.  The uncertainties in the line widths were determined from Monte Carlo random subset sampling, in which $N$ narrow-line subtracted spectra were selected from the list of $N$ spectra, without regard to whether a spectrum had been previously chosen.  The mean and rms spectra were then created from the subset.  The local linear continuum beneath an emission line was set by selecting a random region of width at least 15\,\AA\ from the total allowed continuum region (typically 50\,\AA\ in width, but only 25\,\AA\ for the small region between H$\beta$ and \ion{He}{2}) on either side of the line, and FWHM and $\sigma_{\rm line}$ were then measured and recorded. The process then repeated. This procedure accounts for the effects of including any individual spectrum, with its unique noise properties, in the line width determination.  It further quantifies the uncertainty on the line width from the exact continuum placement.  Distributions of the line width measurements were built up over 1000 realizations, and we report the mean and standard deviation of each distribution as the measurement value and its associated uncertainty. 

Finally, we corrected the emission-line widths for the resolution of the spectrograph following \citet{peterson04}. The measured line width, $\Delta \lambda_{\rm obs}$, may be described as
\begin{equation}
    \Delta \lambda^2_{\rm obs} = \Delta \lambda^2_{\rm true} + \Delta \lambda^2_{\rm disp}
\end{equation}
\noindent where $\Delta \lambda_{\rm true}$ is the intrinsic line width and $\Delta \lambda_{\rm disp}$ is the broadening caused by the spectrograph.  We adopted the FWHM of [\ion{O}{3}] $\lambda 5007$ (13.85\,\AA) as $\Delta \lambda_{\rm obs}$.  The highest-resolution spectrum available for estimating $\Delta \lambda_{\rm true}$ is the STIS G430M spectrum, which has ${\rm FWHM}=6.02$\,\AA\ for [\ion{O}{3}] $\lambda 5007$.  The resolution correction is then $\Delta \lambda_{\rm disp} \approx 12.5$\,\AA.  In Table~\ref{tab:lagwidth} we report the final resolution-corrected emission line widths. 

\section{Black Hole Mass}

The black hole mass is usually determined from reverberation measurements as
\begin{equation}
    M_{\rm BH} = f \frac{c\tau V^2}{G}
\end{equation}
\noindent where $\tau$ is the emission-line time delay, $V$ is the emission line width, $c$ is the speed of light, and $G$ is the gravitational constant.  The factor $f$ is an order-unity scaling factor that accounts for the details of the broad-line region inclination, geometry, and kinematics.  In practice, it is usually not possible to determine the appropriate value of $f$ for each AGN, so a population averaged value, $\langle f \rangle$, is adopted instead.  This "fudge factor" is taken to be the multiplicative factor needed to bring the \msigma\ relationship for AGNs into agreement with the \msigma\ relationship for galaxies with black hole masses determined from dynamical modeling (e.g.,  \citealt{gultekin09,kormendy13,mcconnell13}).  This method should remove any bias from reverberation-based masses on the whole, but results in a factor of $2-3$ uncertainty on the mass of any black hole in particular. Typical values of $\langle f \rangle$ range from 2.8 \citep{graham11} to 5.5 \citep{onken04} in the literature  depending on the exact sample and the analysis methods, with most investigations settling on values of $\sim 4-5$.  Here, we adopt the value of $\langle f \rangle = 4.82$ reported by \citet{batiste17b}, given their careful treatment of morphological effects on measurements of $\sigma_*$ in the \msigma\ relationship.

\citet{peterson04} demonstrated that $\tau_{\rm cent}$ combined with $\sigma_{\rm line}$ produces the least scatter among the predicted masses for NGC\,5548, the AGN with the most independent reverberation experiments (14 separate measurements for H$\beta$ at that time).  We therefore adopt $\tau_{\rm cent}$ for the time delay and $\sigma_{\rm line}$(rms) for the line width of  H$\beta$, and the simple prescription in Equation~3 gives a black hole mass of $M_{\rm BH} = 2.34^{+0.43}_{-0.43} \times 10^7$\,\msun\ for NGC\,3783.  If we instead adopt $\tau_{\rm peak}$ or $\tau_{\rm jav}$ as the H$\beta$ time delay, the predicted mass is consistent within the uncertainties.  If FWHM is adopted instead of $\sigma_{\rm line}$ as the H$\beta$ line width measurement, then a different value for $\langle f \rangle$ must also be adopted (e.g., \citealt{collin06}), but the predicted mass is again consistent within the uncertainties.

The measurements of the other emission lines presented here may also constrain the black hole mass, but they predict a mass that is lower by $\sim 60-70$\%.  However, there is reason to be cautious about this result because the use of the [\ion{O}{3}] doublet to intercalibrate the spectra means that the ability to quantify real variability in spectral features decreases as the wavelength shifts away from the [\ion{O}{3}] lines.  The effects of this can be seen in the noisier light curves for H$\gamma$ and H$\delta$ and the lower peaks for their cross correlation functions. There are no similar strong and unblended narrow emission lines available to improve the spectral calibration of the region near H$\gamma$ and H$\delta$.  This is less of an issue for \ion{He}{2}, which is close in wavelength to H$\beta$ and the [\ion{O}{3}] doublet, however it has its own challenges given the low contrast of the emission line relative to the continuum level.  The different mass predicted by the properties of the  broad \ion{He}{2} line could indicate a physical difference between the parts of the broad line region that are probed by the two emission lines given their different ionization potentials.  If that is the case, then a separate $\langle f \rangle$ would be needed for predicting black hole mass from the \ion{He}{2} emission line.

The black hole mass that we have constrained here is similar to, but slightly smaller than, the mass found by \citet{onken02} of $M_{\rm BH} = 2.9^{+1.1}_{-0.8} \times 10^7$\,\msun, after scaling their H$\beta$ based mass to account for the difference in adopted $\langle f \rangle$ factors.  The improved time sampling in our monitoring program served to decrease the measurement uncertainties, and thus the uncertainties on $M_{\rm BH}$, by $\sim 50$\%.  Further improvement may be possible, but will require moving away from the simplistic mass constraint described above, and instead modeling the full broad-line response as a function of velocity across the emission line profile \citep{pancoast14,grier17,williams18}.  

\begin{figure}
    \epsscale{1.15} 
    \plotone{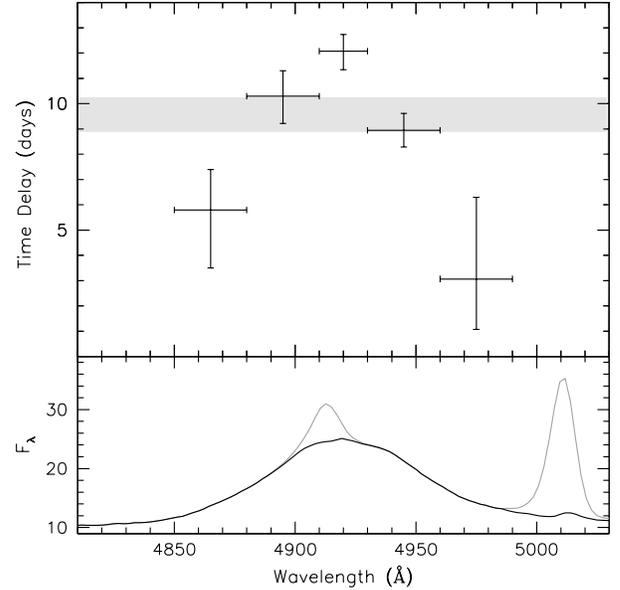}
    \caption{Velocity-resolved time delays (top) across the broad H$\beta$ emission line profile (bottom).  The symmetric response with slightly shorter lags in the red wing of the line is similar to other Seyferts that have been found to have thick disk-like broad line regions with dynamics that are described by a combination of rotation and inflow. }
    \label{fig:velres}
\end{figure}

To investigate whether detailed modeling of the H$\beta$ emission-line response may be successful, we defined five relatively equal-width bins across the wavelength (or velocity) range of the line.  Two bins were defined for each of the red and blue wings, and one bin was defined for the line center.  Light curves for each wavelength bin were then created and their cross-correlation functions were determined, following the same procedures as outlined above for the integrated line emission.  In Figure~\ref{fig:velres} we plot the $\tau_{\rm cent}$ values for all five wavelength bins, with the H$\beta$ line profile displayed in the bottom panel for reference.  The gray horizontal band marks the $\tau_{\rm cent}$ range for the emission line as a whole.  There is a clear velocity-resolved response across the line profile, with a mostly symmetric shape that has shorter time delays in the line wings and a longer time delay in the line center.  A similar, but weaker, signature is also seen in the \ion{He}{2} broad line response, but the less accurate intercalibration of the spectra at the wavelengths of H$\gamma$ and H$\delta$ masks any velocity-resolved response. 

For a physically extended and centrally illuminated BLR, the gas that is closest to the outside observer will exhibit the shortest time delays, while the gas that is farthest from the observer will exhibit the longest time delays.  While the time delay probes the physical arrangement of the BLR gas, the associated velocity relative to the line center probes the kinematics as viewed by the observer (cf.\ \citealt{peterson01} for a full review).  The clear signatures in H$\beta$ and \ion{He}{2} are consistent with the expected velocity-resolved response from a rotating disk of broad line emission, where the time delays measured from redshifted and blueshifted gas would be symmetric around the line center.  The shorter time delay in the longest wavelength (most redshifted) bin of H$\beta$ is similar to what has been seen for a few other AGNs, such as Arp\,151 \citep{bentz10b}, where dynamical modeling of the broad line region has constrained the gas motions to a combination of rotation and inflow \citep{pancoast14}.  Given the clear resolution of different time delays across the H$\beta$ emission line, we expect that modeling of the full reverberation response will provide strong constraints on the BLR geometry and kinematics, and thus the black hole mass.  Modeling of the broad \ion{He}{2} line may allow us to investigate whether there is any evidence for physical differences between the regions of the BLR that are probed by  \ion{He}{2} versus H$\beta$.  Furthermore, upcoming approved MUSE observations with VLT will allow stellar dynamical modeling to constrain the black hole mass in a completely independent way for direct comparison with the reverberation mass.

\section{Summary}

We have carried out a successful photometric and spectroscopic monitoring campaign of NGC\,3783.  Time delays between variations in the continuum and in the broad H$\beta$, H$\gamma$, H$\delta$, and \ion{He}{2} emission lines are clearly detected.  With the reverberation response of the broad H$\beta$ emission line, we constrain a black hole mass of $M_{\rm BH} = 2.34^{+0.43}_{-0.43} \times 10^7$\,\msun.  Clear velocity-resolved signatures across the $H\beta$ profile show a symmetric pattern, with shorter lags in the line wings and a longer lag in the line core.  Future modeling of the full velocity-resolved response will further constrain the black hole mass and the physical details of the broad line region in NGC\,3783.

\acknowledgements

We thank the referee for thoughtful comments that improved the presentation of this work.  MCB gratefully acknowledges support from the NSF through grant AST-2009230 to Georgia State University.  CAO acknowledges support from the Australian Research Council through Discovery Project DP190100252.  MV acknowledges support through NSF grant AST-2009122.  This work makes use of observations from the Las Cumbres Observatory global telescope network.  This research has made use of the NASA/IPAC Extragalactic Database (NED) which is operated by the Jet Propulsion Laboratory, California Institute of Technology, under contract with the National Aeronautics and Space Administration and the SIMBAD database, operated at CDS, Strasbourg, France.

\bibliographystyle{apj}

\clearpage

\end{document}